# Resonance-enhanced spectral funneling in Fabry–Perot resonators with a temporal boundary mirror

Kanghee Lee[1,†], Junho Park[2,†], Seojoo Lee[1], Soojeong Baek[1], Jagang Park[1], Fabian Rotermund,[2,*] and Bumki Min[1,*]

[1]*Department of Mechanical Engineering, Korea Advanced Institute of Science and Technology (KAIST), Daejeon 34141, Republic of Korea*

[2]*Department of Physics, Korea Advanced Institute of Science and Technology (KAIST), Daejeon 34141, Republic of Korea*

[*] Correspondence to: bmin@kaist.ac.kr and rotermund@kaist.ac.kr

[†]These authors contributed equally to this work.

**Abstract**

A temporal boundary refers to a specific time at which the properties of an optical medium are abruptly changed. When light interacts with the temporal boundary, its spectral content can be redistributed due to the breaking of continuous time-translational symmetry of the medium where light resides. In this work, we use this principle to demonstrate, at terahertz (THz) frequencies, the resonance-enhanced spectral funneling of light coupled to a Fabry–Perot resonator with a temporal boundary mirror. To produce a temporal boundary effect, we abruptly increase the reflectance of a mirror constituting the Fabry–Perot resonator and, correspondingly, its quality factor in a step-like manner. The abrupt increase in the mirror reflectance leads to a trimming of the coupled THz pulse that causes the pulse to broaden in the spectral domain. Through this dynamic resonant process, the spectral contents of the input THz pulse are redistributed into the modal frequencies of the high-$Q$ Fabry–Perot resonator formed after the temporal boundary. An energy conversion efficiency of up to 33% was recorded for funneling into the fundamental mode with a Fabry–Perot resonator exhibiting a sudden Q-factor change from 4.8 to 48. We anticipate that the proposed resonance-enhanced spectral funneling technique could be further utilized in the development of efficient mechanically tunable narrowband terahertz sources for diverse applications.

# 1. Introduction

By harnessing temporal degrees of freedom, time-variant photonic platforms have enabled diverse optical functionalities, such as spectral conversion [1-28], nonreciprocal transmission [29-31], topologically nontrivial phases [32], synthetic dimensions [33,34], and beam steering in reflection and refraction [35-37]. Among these, spectral conversion of light in time-variant media has a long history of theoretical research dated back to the 1950s [1-5]. The key concept underlying the spectral conversion of light was experimentally verified in earlier measurements performed with suddenly created plasma [6-10]. Later, with the functional advancement of optical fibers and devices, spectral conversion of light was demonstrated in various time-variant photonic platforms, such as optical fibers [11-13], waveguides [14-16], microresonators [17-19], and metamaterials [20-26]. It is also interesting to note that spectral conversion is found to occur even at a few-photon level [12-14], illustrating that, for a certain type of application, time-variant photonic platforms can be considered an alternative to traditional nonlinearity-based frequency converting devices. While various time-variant platforms with sophisticated designs can be employed, one of the simplest examples is the spectral conversion of light achieved by its interaction with a step-like temporal variation of the constituting element in photonic platforms. Analogous to the spatial boundary (or interface) existing between two media of distinct static optical properties, the time at which a step-like variation in optical properties occurs is referred to as a temporal boundary. Due to the broken time-translation symmetry of the medium in which light resides, the frequency of light (or the energy of a photon) can be changed in the presence of the temporal boundary [1, 7-10, 21]. Keeping this basic principle in mind, one can devise various intriguing schemes for the spectral redistribution of light. For example, in our previous work, we proved that spectral conversion can be observed in a time-variant two-dimensional metasurface, of which the effective surface conductivity abruptly changes at the temporal boundary [24]. However, due to the limited interaction time between the pulse of light and the temporal boundary, the energy conversion efficiency was found to be relatively low, and the amplitude transmission was slightly greater than unity. It is also worthwhile to note that, to the best of our knowledge, most time variant resonant platforms employed for frequency shift or spectral conversion of light have induced an additional loss, thereby leading to inevitable broadening of the spectrum of converted or shifted output compared to that of the input.

In this work, we propose a resonant platform that can be rapidly changed from a low Q to a high Q state for resonance-enhanced spectral conversion of light. The decrease in resonator loss enables *funneling* of spectral components of the input into a narrower resonance formed posterior to the temporal boundary. More specifically, we experimentally demonstrate that resonance-enhanced spectral funneling can be observed in a Fabry–Perot (FP) resonator consisting of a spatial and temporal boundary mirror. Resonance-enhanced spectral funneling is a dynamic process where the input pulse interacts with the spatial and temporal boundaries of the FP resonator. Consequently, a proper spatiotemporal design strategy is requested for the optimization of the resonance-enhanced spectral funneling process. We show that this optimization task can be accomplished by adjusting the (spatial) length of the FP resonator and the time delay between the input pulse and the temporal boundary. While the energy conversion efficiency depends on the relative amount of spectral shift, we show that the efficiency can exceed 30% (for the relative amount of spectral shift of 0.26), which is orders of magnitude higher than the value achieved in our previous THz metasurface platform [24] and is comparable to the value observed in a THz waveguide system [16]. All these experimental observations are verified by comparison with theoretical calculations.

## 2. Results and discussion

### 2.1 Implementation of FP resonators with a temporal boundary mirror

To experimentally verify the proposed concept, we constructed a THz FP resonator consisting of two different types of mirrors: a temporal boundary mirror and a time-invariant mirror. For ease of reference, these two mirrors will be termed hereafter as the temporal and static mirrors, respectively. The temporal mirror was implemented by using a semi-insulating gallium arsenide (GaAs) substrate, the reflectance of which can be increased abruptly by ultrafast optical pumping [38]. The formation of a surface conductive layer by pumping occurs at a time scale of approximately less than 100 fs [27,28,39]. Considering the carrier frequency of the incident THz pulse, the rising time for the formation of a conductive layer can be regarded as being relatively abrupt. In Section 2.6, we will discuss the role of this abruptness on the resonance-enhanced spectral funneling process in detail. In the experiment, a pump pulse at a center wavelength of 800 nm was incident on the GaAs substrate (with a fluence of approximately 100 µJ/cm$^2$). Once excited, the reduced transmission through the GaAs substrate was maintained for at least 100 ps with a very slow rate of recovery (see Fig. S1 of supplementary material for characterization of the temporal mirror). The reduction in transmission is attributed to the creation of a photoconductive layer on the GaAs surface and can also be quantified by a sudden increase in the GaAs refractive index (or the effective surface conductivity). The refractive index of semi-insulating GaAs is approximately 3.6 in the frequency range of interest [38]. Assuming that a photoconductive layer is created on the surface of the GaAs substrate with a penetration depth of approximately 1 µm [40], the refractive index of the photoconductive GaAs layer is estimated to be at least 50 in the experiments. The static mirror was constructed by patterning and depositing an array of gold wires on a 1 µm-thick polyimide film (see Fig. S2 of the supplementary material for characterization of the static mirror). The distance between neighboring wires was set to 40 μm, and the thickness and width of each wire were designed to be 200 nm and 4 μm, respectively. When the polarization of a THz pulse is parallel to the wire direction, the array of wires can work as a partial reflector (see Fig. S2). While the position of the temporal mirror was fixed in the experiment, the position of the static mirror was adjusted with a motorized actuator enabling precise control of the cavity length. FP resonators with cavity lengths of 120, 250 and 360 μm exhibited fundamental mode $Q$-factors of approximately 3.9, 4.8 and 5.1 before the temporal boundary and 14.3, 48 and 73 after the boundary. The decrease in the fundamental mode $Q$-factor for a shorter cavity length is attributed to the frequency-dependent reflectance of the static mirror (see Fig. S2).

## 2.2 Characterization of resonance-enhanced spectral funneling

For characterization of resonance-enhanced spectral funneling, we employed an ultrafast THz time-domain spectroscopy setup constructed by using a Ti:sapphire regenerative amplifier laser system (Spitfire Ace, Spectra-Physics). In the first series of measurements, a single-cycle THz pulse was transmitted through the temporal mirror in the low reflectance state (or the low $Q$-state of the FP resonator) so that part of the incident THz pulse could be coupled to the FP resonator (Fig. 1a). As the THz pulse was coupled, the optical pump pulse abruptly increased the reflectance of the temporal mirror (Fig. 1b). As shown in Fig. 1b, the pump pulse was launched into the FP resonator from the static mirror side. Then, the coupled THz pulse was trapped in the resonator before eventually leaking through the static mirror (Fig. 1c). Note that the reflectance of the static mirror was much lower than that of the temporal mirror posterior to the temporal boundary; therefore, the output spectrum was measured from the static mirror side. Through this process, the single-cycle input THz pulse was transformed into a multicycle transmitted pulse consisting of major spectral components determined mostly by the modal frequencies of the FP resonator (see Fig. 2a and 2b). In this measurement, the cavity length of the FP resonator was set to approximately 250 μm.

The spectral funneling process was found to be dependent on the arrival time of the ultrafast pump pulse (or the time delay). More interestingly, the field amplitudes at the fundamental and second-order modal frequencies (drawn with a red and blue line, respectively) of the FP resonator were maximized at distinct values of the time delay (Fig. 2a and 2b). These optimized timing conditions can be found by measuring a transmitted spectral amplitude as a function of the time delay (Fig. 2c). The difference in the optimized time delays for fundamental and second-order resonance-enhanced funneling can be qualitatively explained based on the degree of spectral overlap; when the overlap between the spectrum of the input THz pulse and the modal frequency of interest is relatively large, the funneling effect can be resonantly enhanced by maximizing the portion of the input THz pulse coupled into the FP resonator. On the other hand, when the spectral overlap is relatively small, a large amount of spectral broadening and correspondingly a sharp trimming of the coupled THz pulse are required. For example, to maximize funneling into the second- and third-order modes, a temporal boundary should be created at the instance when the peaks of the THz pulse are transmitted through the GaAs/air interface (or the temporal mirror). The measured time-delay-dependent transmitted spectral amplitude shown in Fig. 2c supports the aforementioned argument. As seen in the right panel of Fig. 2c, the transmitted field amplitudes at the second- and third-order modal frequencies undulate with the input THz pulse waveform. As a side note, two extreme cases need to be mentioned. When the input THz pulse is transmitted through the FP resonator before the temporal boundary, the input THz pulse is filtered by the low Q FP resonator formed before the temporal boundary (see the plots drawn with green lines in Fig. 2a and 2b). In contrast, when the input THz pulse arrives at the FP resonator after the temporal boundary, the input THz pulse can barely be transmitted through the FP resonator (see the plots drawn with purple lines in Fig. 2a and 2b).

As the spectrum of the input THz pulse and that of the transmitted pulse can overlap, the energy conversion efficiency should be carefully defined. In the following discussion, the energy conversion efficiency $\eta$ is defined as

$$\eta = \frac{\int (|E_t^2(\omega)| - |E_i^2(\omega)|) d\omega}{\int |E_i^2(\omega)| d\omega}, \tag{1}$$

where $E_i(\omega)$ and $E_t(\omega)$ denote the spectral amplitudes of the input and transmitted pulses, respectively. Throughout the manuscript, the reference input THz pulse is taken as that directly transmitted through the semi-insulating GaAs substrate without pulsed excitation. In calculating the numerator, the integral is taken over the frequency interval near the resonance of interest (more specifically, only when the transmitted spectral amplitude is larger than the input). Figure 2d graphically illustrates how the energy conversion efficiency is evaluated for a specific case of resonance-enhanced spectral funneling into the fundamental mode of the FP resonator.

### 2.3 Resonance-enhanced spectral funneling: single-cycle pulse excitation

Figure 3a shows the resonance-enhanced spectral funneling, which was experimentally measured for the FP resonator with a variation in the cavity length (from the top panel, 360 μm, 250 μm, 180 μm, 140 μm, and 120 μm). In each of the measurements, we separately optimized the time delay to maximize spectral components funneled to the fundamental (drawn with red lines) or the second-order (drawn with blue lines) resonance frequencies of the FP resonator. It is worthwhile to note that the funneled spectral components exceed the spectral amplitudes of the input THz pulse at the resonance frequency. The energy conversion efficiencies were estimated (for the fundamental resonance mode) to be 18% for the FP resonator with $L = 360$ μm, 33% for $L = 250$ μm, 27% for $L = 180$ μm, 11% for $L = 140$ μm, and 5% for $L = 120$ μm. In addition, the conversion efficiencies into the second-order resonance mode were found to be 11% for $L = 360$ μm, 0.9% for $L = 250$ μm, and 0.2% for $L = 180$ μm. Due to the limitation in the acquisition time window, it was not possible to record fully transmitted THz pulse waveforms, especially for the FP resonator with longer cavity length (i.e., equivalently here, a longer cavity lifetime). The conversion efficiency was estimated based on the measured portion of the transmitted waveform. Nonetheless, the measured energy conversion efficiencies were much higher than those reported for nonlinear spectral conversion processes [41-44]. Recently, a highly efficient spectral conversion process with up to 23% efficiency in a time-variant THz waveguide system was reported [16]; the conversion efficiencies reported here were found to be comparable to those obtained from the time-variant THz waveguide system (under similar experimental conditions). From the measurements, we are able to provide two important key observations. First, a sudden increase in the reflectance of the temporal mirror leads to a trimming of the coupled THz pulse waveform in the time domain and the spectral broadening of the transmitted THz pulse beyond the extent of its input spectrum [27,28]. Because of this broadening, the spectral contents of the input THz pulse can be redistributed to various higher or lower FP resonator modes and, especially, to the modes not overlapping with the spectrum of the incident THz pulse. Second, after the temporal boundary, funneled spectral components at modal frequencies are enhanced due to the increased resonator $Q$-factors; therefore, part of the modes are radiated through the static mirror with a much longer decay time. As a result, resonance-enhanced spectrally funneled peaks, even larger in amplitude than the components in the input spectrum, are observable. This enhancement illustrates that resonance-enhanced spectral funneling is not passive filtering but a dynamic process that involves active spectral conversion.

**2.4 Numerical analysis**

To qualitatively corroborate our experimental observation and gain a more in-depth understanding of the underlying mechanisms, we employed temporal coupled mode theory (TCMT) and phenomenologically investigated the resonance-enhanced spectral funneling process [45-47]. In particular, we considered a specific FP resonator with $n$ resonance modes connected to two ports (ports 1 and 2). Then, the mode amplitude $\boldsymbol{a}(t)$ is related to the incoming and outgoing waves $\boldsymbol{s}_+(t)$ and $\boldsymbol{s}_-(t)$ by the following governing equations:

$$\frac{d\boldsymbol{a}(t)}{dt} = j\boldsymbol{\Omega}\boldsymbol{a}(t) - (\boldsymbol{\Gamma}_{\mathbf{ex}} + \boldsymbol{\Gamma}_{\mathbf{in}})\boldsymbol{a}(t) + \boldsymbol{K}^T\boldsymbol{s}_+(t), \tag{1}$$

$$\boldsymbol{s}_-(t) = \boldsymbol{C}\boldsymbol{s}_+(t) + \boldsymbol{K}\boldsymbol{a}(t), \tag{2}$$

where $\boldsymbol{\Omega}$, $\boldsymbol{\Gamma}_{\mathbf{ex}}$ and $\boldsymbol{\Gamma}_{\mathbf{in}}$ are $n \times n$ Hermitian matrices describing the resonance frequencies and radiative and intrinsic loss of the resonator, respectively, while $\boldsymbol{K}$ and $\boldsymbol{C}$ are a $2 \times n$ coupling matrix and a $2 \times 2$ scattering matrix, respectively. Here, the column vectors $\boldsymbol{s}_{\pm}(t)$ have two components $s_{\pm 1}(t)$ and $s_{\pm 2}(t)$ that represent incoming (outgoing) waves to (from) ports 1 and 2. The coupling matrix $\boldsymbol{K}$ can be arranged as $\boldsymbol{K}^T = [\boldsymbol{k}_1 \quad \boldsymbol{k}_2]$, where the column vectors $\boldsymbol{k}_1 = [k_{11}, k_{12}, .., k_{1m}, .., k_{1n}]^T$ and $\boldsymbol{k}_2 = [k_{21}, k_{22}, .., k_{2m}, .., k_{2n}]^T$ describe the coupling between $\boldsymbol{a}(t)$ and $\boldsymbol{s}_-(t)$ through the temporal and static mirrors, respectively. Considering our specific experimental scheme in which the incoming wave is from port 1, $s_{+1}(t)$, is related to the electric field of the input THz pulse, while the incoming wave from port 2, $s_{+2}(t)$, is set to zero. Then, we can numerically calculate the mode amplitude $\boldsymbol{a}(t)$ and the corresponding outgoing (transmitted) THz wave from port 2, $s_{-2}(t)$. In reproducing the measured data through the application of the TCMT, the components of $\boldsymbol{k}_1$ and $\boldsymbol{k}_2$ were assumed to have amplitudes that were proportional to the amplitude transmission through the temporal and static mirror. For the assignment of the phase to each of the coupling vector components, the modal field symmetry was considered. Subsequently, the matrix elements of $\boldsymbol{\Gamma}_{\mathbf{ex}}$ were obtained from the relation $\boldsymbol{K}^\dagger \boldsymbol{K} = 2\boldsymbol{\Gamma}_{\mathbf{ex}}$ [45-47]. In addition, the matrix $\boldsymbol{\Gamma}_{\mathbf{in}}$ was estimated by considering the surface conductivity of the static mirror at the modal frequencies of the FP resonator (see Fig. S2 in the supplementary material). In estimating the matrix elements of $\boldsymbol{\Gamma}_{\mathbf{in}}$, we assumed that the lumped loss induced by the temporal mirror was negligible compared to the intrinsic loss of the static mirror.

Based on the aforementioned procedure, we calculated the spectral amplitude of the transmitted pulse through the FP resonator with the assumption of an abruptly changing coupling vector, $\boldsymbol{k}_1(t)$, of the temporal mirror. Specifically, in the calculations, each component of the column vector $\boldsymbol{k}_1(t)$ was decreased in its amplitude by the same amount, which can be justified by the fact that the amplitude transmission through the temporal mirror was slightly dispersive (see Fig. S1 of supplementary material). These calculated spectral amplitudes are plotted in the panels of Fig. 3b, where each plot in the vertical panels corresponds to the measured spectral amplitude in each of the panels in Fig. 3a. To simulate a more realistic gradual development of the temporal boundary, the time-varying behavior of $k_{1m}(t)$ was modeled as follows:

$$|k_{1m}(t)|^2 = \left|k_{1m}^{(i)}\right|^2 \frac{1}{\pi}\int_{(t-t_p)/\tau}^{\infty} e^{-x^2}dx + \left|k_{1m}^{(f)}\right|^2 \left(1 - \frac{1}{\pi}\int_{(t-t_p)/\tau}^{\infty} e^{-x^2}dx\right), \quad (3)$$

where $k_{1m}^{(i)}$ and $k_{1m}^{(f)}$ denote the asymptotic coupling components far before and after the temporal boundary, respectively. In this expression, the excitation time of the pump pulse and the characteristic transition time of the temporal mirror are denoted by $t_\mathrm{p}$ and $\tau$, respectively. For calculation of the spectral amplitudes shown in Fig. 3b, we assumed the characteristic transition time $\tau$ of 150 fs by considering the actual pump pulse width. Here, several remarks should be made on the numerical analysis. First, the TCMT is known to not be applicable in a strict sense for describing low-Q resonance modes. This raises the question of whether the FP resonator prior to the temporal boundary can be properly analyzed by the TCMT [45-47]. Nonetheless, we found that the TCMT calculation result shows excellent qualitative agreement with the measured data. We attribute this level of consistency in part to the fact that the input THz pulse interacts with this low-$Q$ state only in the relatively short period of cavity loading time. Second, the spectral amplitude peaks measured for the FP resonators with cavity lengths of 360 μm and 250 μm (shown in the first and second panels from the top in Fig. 3a) are broader than those calculated from the TCMT calculations (shown in the first and second panels of Fig. 3b). These broadened peaks in the measured spectral amplitudes are attributed to the limited spectral resolution of the THz detection scheme employed in our setup. Finally, the resonance-enhanced spectral funneling can also be analyzed with the virtual field method, as done in our previous work on time-variant metasurfaces [24], but, for simplicity, the details will not be discussed here.

## 2.5 Resonance-enhanced spectral funneling: multicycle pulse excitation

To further understand the resonance-enhanced spectral funneling process, we conducted additional experiments using a multicycle input pulse with a center frequency of 0.6 THz. In these measurements, the cavity length of the FP resonator was adjusted to approximately 360 μm, and a multicycle pulse with a pulse width of approximately 10 ps was prepared by transmitting a single-cycle input pulse through cascaded bandpass filters [24,44]. As shown in Fig. 4a, the spectral content of the multicycle input pulse is redistributed and enhanced at the fundamental resonance frequency of 0.42 THz as well as at higher-order resonance frequencies, such as 0.84 THz and 1.25 THz (Fig. 4a). The measured energy conversion efficiencies were 2.6% for resonance-enhanced funneling into the fundamental mode and 4.3% for the second-order resonance mode at 0.8 THz. These efficiencies are found to be lower than the corresponding values (18% for the fundamental mode and 11% for the second-order mode) measured with a single-cycle pulse. This observed efficiency drop is attributed to the lower utilization of a multicycle input pulse in the funneling process. Considering a cavity lifetime of ~1.8 ps of the FP resonator prior to the temporal boundary, the multicycle input pulse (of a width of ~10 ps) can be coupled less effectively to the resonator than the single-cycle input pulse (of a width of ~2 ps), leading to its decreased interaction with the temporal boundary. Our preliminary analysis suggests that in the limiting case where the pulse width becomes substantially larger than the cavity lifetime, the energy conversion efficiency scales linearly with the inverse pulse width. Predicted by numerical analyses, this tendency matches well with the observed efficiency drop measured in the experiment with a multicycle THz pulse.

Notably, the phase of the resonance-enhanced funneled component can be fully controlled by adjusting the time delay between the input THz and the optical pump pulses. To show the phase controllability, the resonance-enhanced funneled components at the fundamental and second-order resonance frequencies are plotted in the complex plane with a variation in the time delay (Fig. 4b and c). As shown in the figure, the field trajectory in the complex plane evolves with respect to the time delay and distinctively depends on the sign of $\omega - \omega_i$, where $\omega$ and $\omega_i$ denote the funneling and input (angular) frequencies, respectively. Although qualitatively similar tendencies were also reported for spectral conversion with a time-variant metasurface [24], it should be emphasized that in this work, both the frequency and phase of the funneling components could be controlled in real time by adjusting the cavity length and the time delay. All these characteristics can be confirmed by comparing the measured field trajectories with those calculated from the TCMT (Fig. 4b and c). Due to limitations in our setup, an FP resonator with a much longer cavity length could not be experimentally characterized. However, a numerical calculation suggests the existence of a tradeoff relation between the spectral shift and the conversion efficiency. Figure 4d shows the calculated spectral amplitude for an FP resonator with a length of 6 mm (with a fundamental resonance frequency of 0.025 THz). As predicted, the input spectrum is redistributed, and resonance is enhanced at the densely spaced modal frequencies of the longer FP resonator. We note that the funneled spectral amplitude is inversely proportional to the difference between the input and funneled mode frequencies. This tradeoff is attributed to the spectral broadening induced by the temporal boundary [24].

## 2.6. The role of temporal abruptness in resonance-enhanced spectral funneling

In a realistic situation, the temporal mirror is realized with a finite characteristic transition time $\tau$, as has been considered in the above numerical calculations. It can be predicted that the spectral broadening and the spectral funneling depend on the characteristic transition time. Figure 5 shows the spectral amplitudes at the fundamental and second-order resonance frequencies plotted as a function of the characteristic transition time. For this calculation, we consider the transmission of a single-cycle pulse through the time-variant FP resonator with a cavity length of 250 μm (also used for calculating the plot in Fig. 3b). With an increase in the characteristic transition time from 50 fs to 2 ps, the spectral amplitude funneled into the second-order mode is substantially reduced, while the amplitude funneled into the fundamental mode is gradually decreased. The distinct scaling behaviors of funneling efficiencies are associated with the balancing between spectral broadening and resonance enhancement. For the specific situation shown in Fig. 5, a substantial amount of spectral broadening by sharp trimming of the coupled THz pulse is required for efficient funneling into the second-order modal frequency. On the other hand, due to a large degree of spectral overlap, sharp trimming plays a lesser role in the observation of efficient funneling into the fundamental modal frequency. As the increase in the transition time sharply reduces spectral broadening, spectral funneling into the second-order modal frequency is more severely affected than the funneling into the fundamental frequency. This discrepancy also implies that for certain types of applications requiring a small amount of spectral shift, one could utilize a slower modulation technique, such as electrical modulation in the microwave range or pulsed excitation at optical frequencies.

## 3. Conclusion

In this work, we propose the use of a temporal boundary mirror in the construction of a FP resonator for the resonance-enhanced spectral funneling of an incident pulse. The method is based on a simple conceptual idea; first, the incident pulse is efficiently launched into the FP resonator in its initial low $Q$-state. Then, the abrupt increase in the mirror reflectance leads to trimming of the coupled pulse waveform and, consequently, its broadening in the spectral domain. Through this process, the input pulse is transformed into a transmitted pulse consisting of spectral components tightly confined to the modal frequencies of the FP resonator formed after the temporal boundary. Notably, the resonance-enhanced spectral funneling process can be utilized to produce a peak in the spectral domain, where the spectrum of the input pulse minimally overlaps. In contrast to previous resonant time-variant platforms, the proposed method relies on a sudden loss decreasing mechanism in the resonator, thereby leading to efficient spectral funneling into a narrower resonance mode formed after the temporal boundary. This spectral shifting, focusing and enhancing capability can be further utilized to implement a bright narrowband THz source, which is of vital importance in ultrafast spectroscopies [43, 48-50]. In addition, the proposed technique could be utilized for an efficient variable frequency synthesizer, thereby contributing to the development of functional devices for future THz communications [51-54]. Last but not least, a similar resonance-enhanced spectral funneling process might be observed at optical frequencies by utilizing materials that exhibit a large and rapid change in properties in the optical domain [55-58].

**Figure captions**

**Figure 1 Schematic illustration of the interaction between the input pulse and the temporal boundary.** (a) Prior to the temporal boundary, $t_{pump}$, an input THz pulse is launched into the FP resonator through the temporal mirror (or the GaAs substrate). (b) When the THz pulse is coupled to the FP resonator, the surface of the GaAs substrate becomes abruptly conductive by ultrafast optical excitation. (c) After the temporal boundary, the trapped THz pulse leaks gradually through the static mirror (or the patterned film) with its major spectral components matching the modal frequencies of the FP resonator.

**Figure 2 Transmitted THz waveforms and corresponding spectral amplitudes and intensities through the FP resonator with a cavity length of approximately 250 μm.** (a) Time traces of the input THz pulse (top panel), transmitted THz pulses with optimized time delays for fundamental (middle panel, red line) and second-order (middle panel, blue line) modal frequencies, and reference pulses (bottom panel, green and purple lines). The green line corresponds to the transmitted THz waveform measured without pulsed excitation, while the purple line corresponds to the transmitted THz waveform measured with pulsed excitation approximately 10 ps prior to the arrival of an input THz pulse on the GaAs substrate. (b) Spectral amplitudes of the corresponding transmitted THz pulses shown in (a). (c) Spectral amplitudes plotted as a function of the time delay. Optimized time delays for fundamental and second-order modal frequencies are denoted by white dashed lines. (d) Graphical representation of the energy conversion efficiency estimation. The reference input THz pulse is taken as that directly transmitted through the semi-insulating GaAs substrate without pulsed excitation.

**Figure 3 Measured and calculated spectral amplitudes for FP resonators with different cavity lengths.** (a) Measured spectral amplitudes transmitted through the FP resonator with different cavity lengths (from the top panel, 360 μm, 250 μm, 180 μm, 140 μm, and 120 μm). In each panel, measured spectral amplitudes optimized for the fundamental (red lines) and second-order (blue lines) spectral components are plotted along with the spectral amplitude of the input THz pulse (black lines). In these measurements, a single-cycle input pulse is launched into the FP resonator. (b) Calculated spectral amplitudes transmitted through the FP resonator with different cavity lengths (from the top panel, 360 μm, 250 μm, 180 μm, 140 μm, and 120 μm). (c) Calculated map of spectral amplitudes with pump delays optimized for maximizing the component funneled to the fundamental resonance frequency. (d) Calculated map of spectral amplitudes with pump delays optimized for maximizing the component funneled to the second-order resonance frequency.

**Figure 4 Resonance-enhanced spectral conversion with a multicycle input pulse.** (a) Measured spectral amplitudes transmitted through the FP resonator with pump delays optimized for maximizing the component funneled to the fundamental (red line) and second-order (blue line) resonance frequencies. In these measurements, a multicycle input pulse is launched into the FP resonator. (b) Complex amplitude trajectory of the funneled component at 0.42 THz plotted as a function of the pump delay. Measured complex amplitudes are drawn with red circles, while the calculated complex amplitudes are drawn with a red line. Note that the funneled frequency is lower than the peak frequency of the input THz pulse. (c) Complex amplitude trajectory of the funneled component at 0.85 THz plotted as a function of the pump delay. Measured complex amplitudes are drawn with blue squares, while the calculated complex amplitudes are drawn with a blue line. Note that the funneled frequency is higher than the peak frequency of the input THz pulse. (d) Calculated spectral amplitudes (drawn with a red line) transmitted through the FP resonator with a cavity length of 6 mm. The spectral amplitude of the input THz pulse is drawn with a black line along with a fitting line $|\omega - \omega_i|^{-1}$. Here, $\omega_i$ denotes the center (angular) frequency of the input pulse.

**Figure 5 Role of the characteristic transition time in resonance-enhanced spectral funneling.** Spectral amplitudes at the fundamental (red color) and second-order (drawn with red color) resonance frequencies plotted as a function of the characteristic transition time. The inset shows the exemplary time-varying behavior of $|k_{1i}|^2$ plotted by assuming the characteristic time of 1 ps.

**Research funding**

This work was supported by National Research Foundation of Korea (NRF) through the government of Korea (NRF-2017R1A2B3012364 and NRF-2019R1A2C3003504). The work was also supported by the center for Advanced Meta-Materials (CAMM) funded by Korea Government (MSIP) as Global Frontier Project (NRF-2014M3A6B3063709). K.L. acknowledges support from NRF (grant no. NRF-2020R1C1C1009098). S.L. was supported by an NRF grant funded by the Korean government (MSIT) (grant no. NRF-2020R1C1C1012138)

**Conflict of interest statement**

The authors declare no conflict of interest regarding this article.

**Figures**

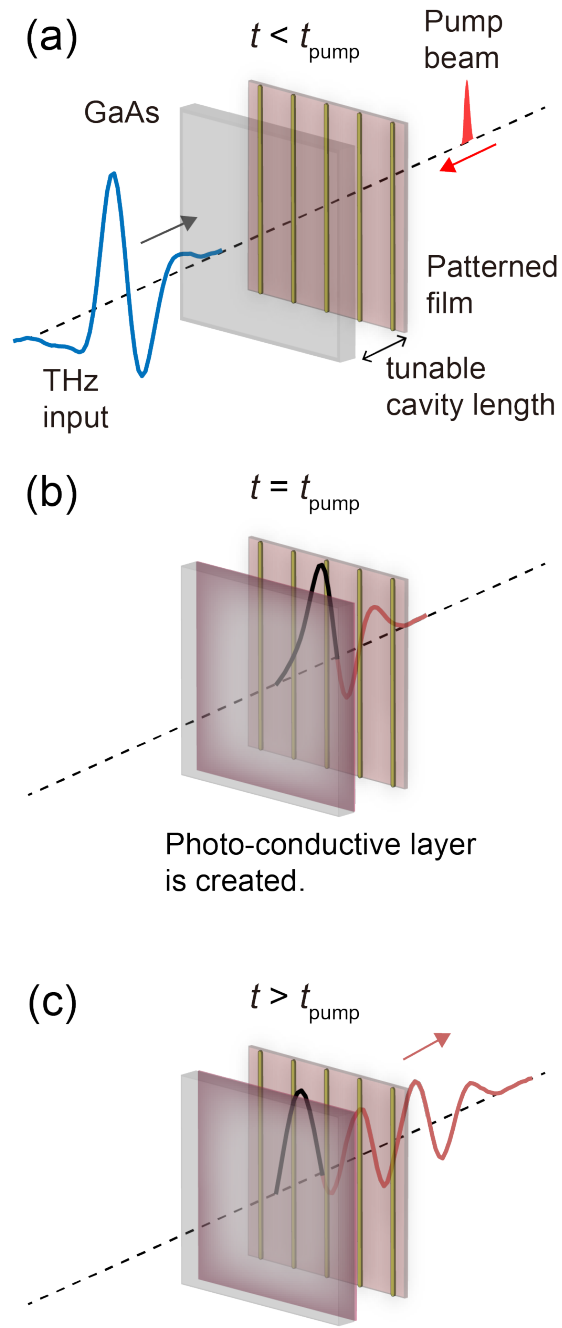

Fig. 1 K. Lee et al.

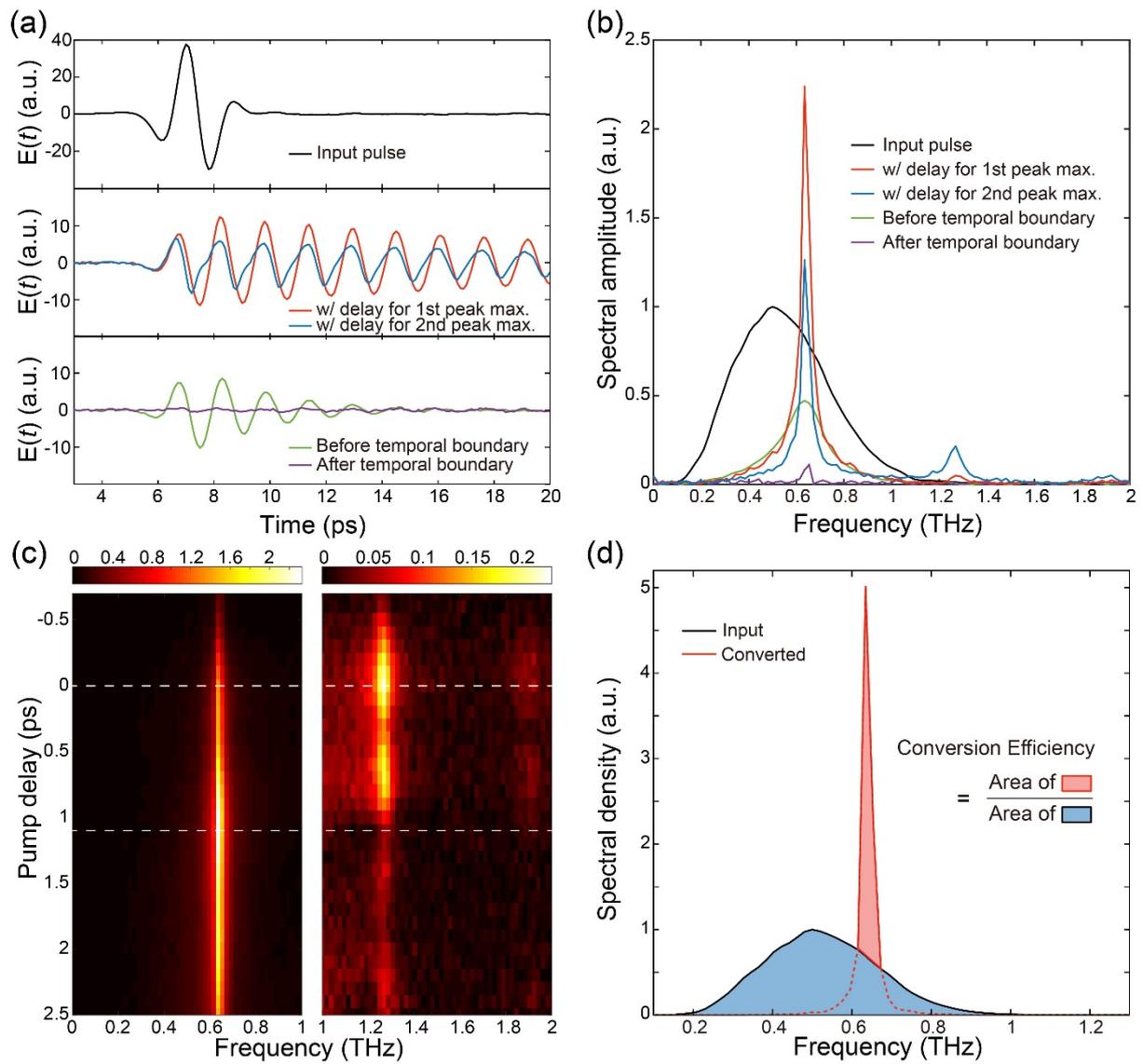

Fig. 2 K. Lee et al.

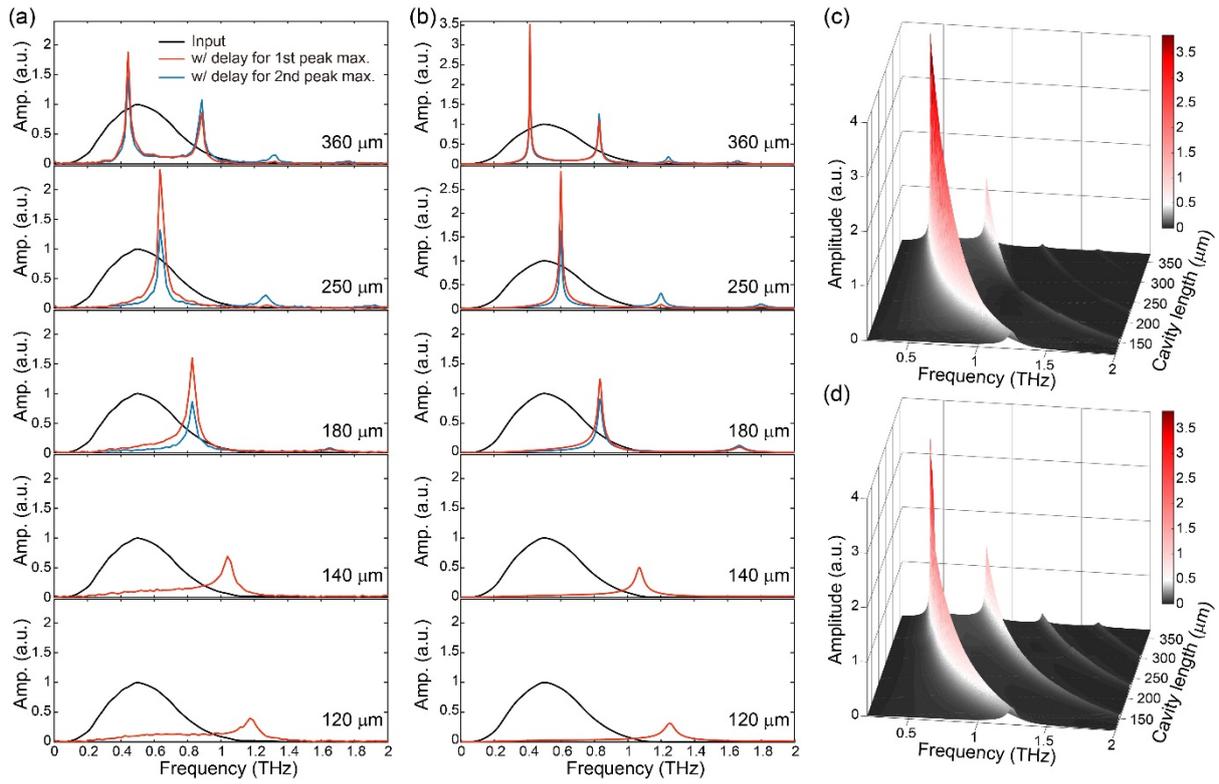

Fig. 3 K. Lee et al.

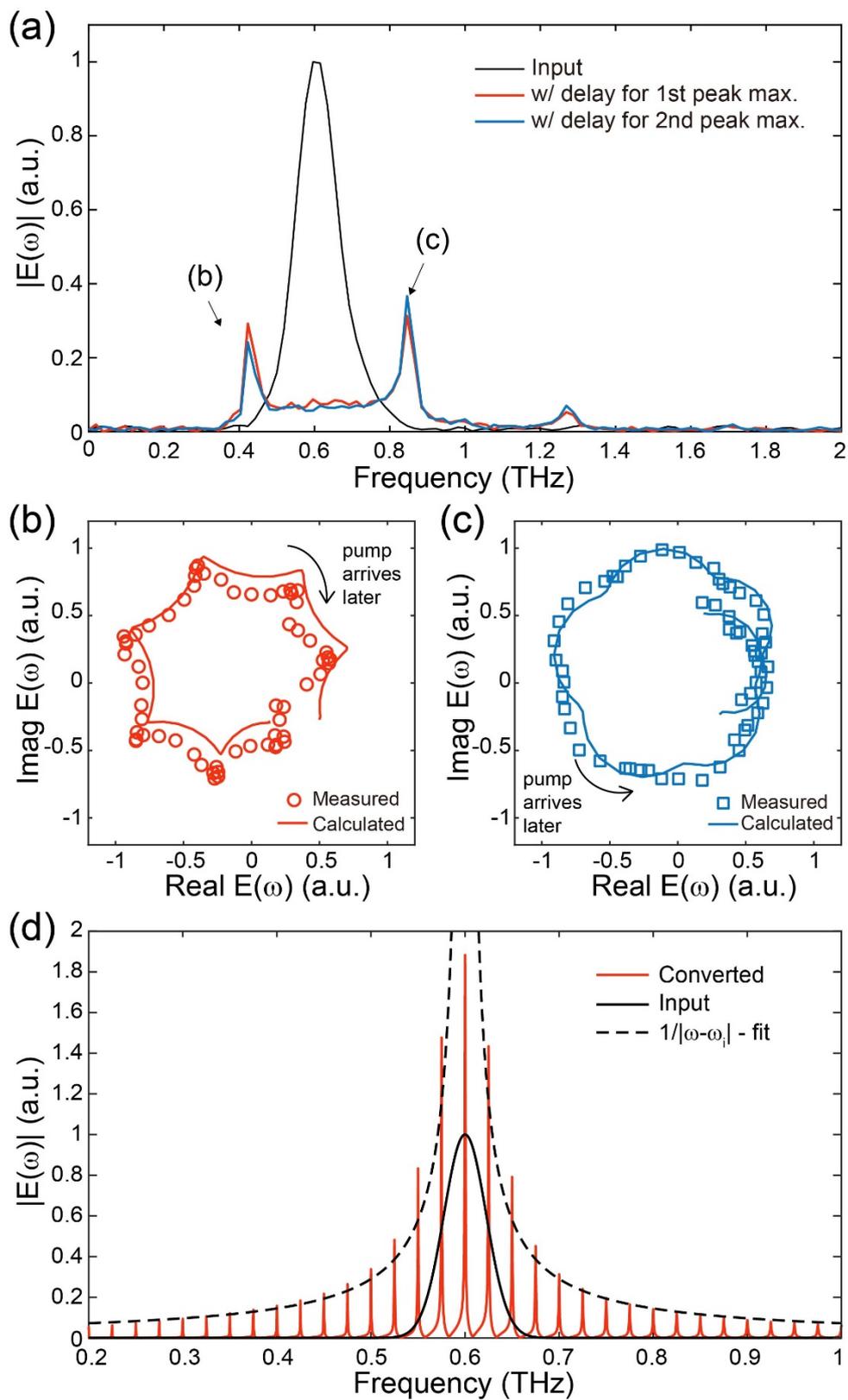

Fig. 4 K. Lee et al.

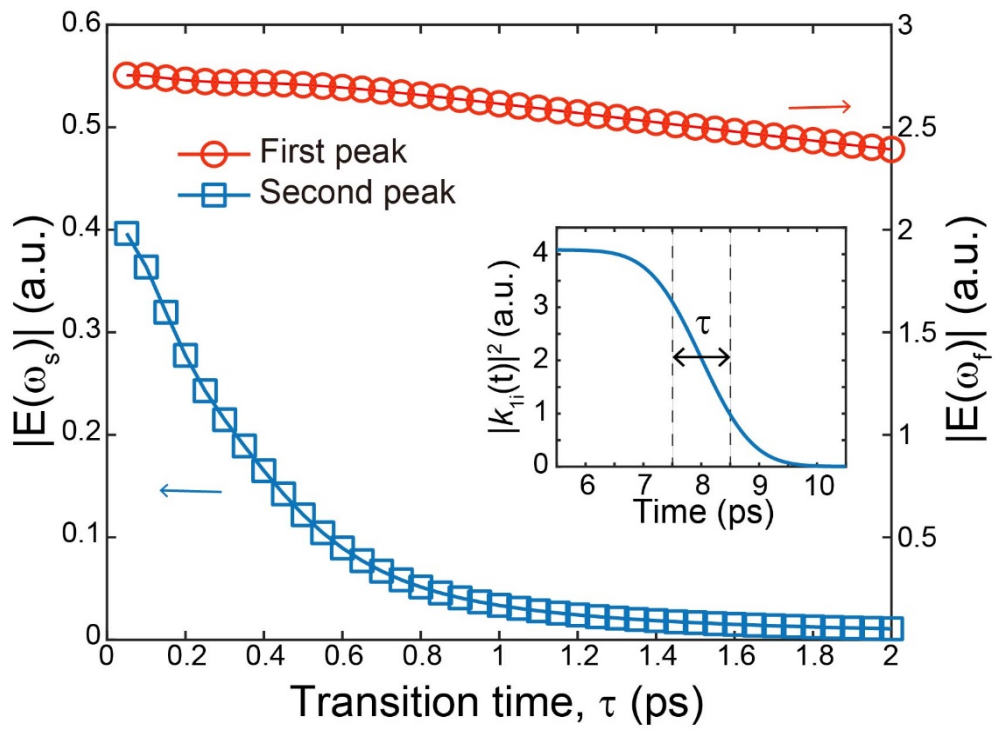

Fig. 5 K. Lee et al.

# Supplementary material: Resonance-enhanced spectral funneling in Fabry–Perot resonators with a temporal boundary mirror

**1. Characterization of the temporal mirror**

The temporal and spectral transmission properties of the temporal mirror (i.e., an optically pumped GaAs substrate) are shown in Fig. S1. To illustrate the step-like time-varying behavior of the temporal mirror upon ultrafast excitation, we plot the peak field of the transmitted THz pulse through the temporal mirror as a function of the pump delay in Fig. S1a (here, the positive pump delay corresponds to the situation after the temporal boundary). As seen in the plot, the transmission reduced by the pump pulse excitation was maintained up to a pump delay of 100 ps with a very slow recovery rate. Figure S1b shows the amplitude transmission spectra through the temporal mirror after the temporal boundary (measured at three different pump delays, 4 ps, 50 ps and 100 ps). The measured amplitude transmission spectra show slight dispersiveness in the frequency range of interest [1], which justifies the assumption made regarding the temporal mirror in the TCMT calculations.

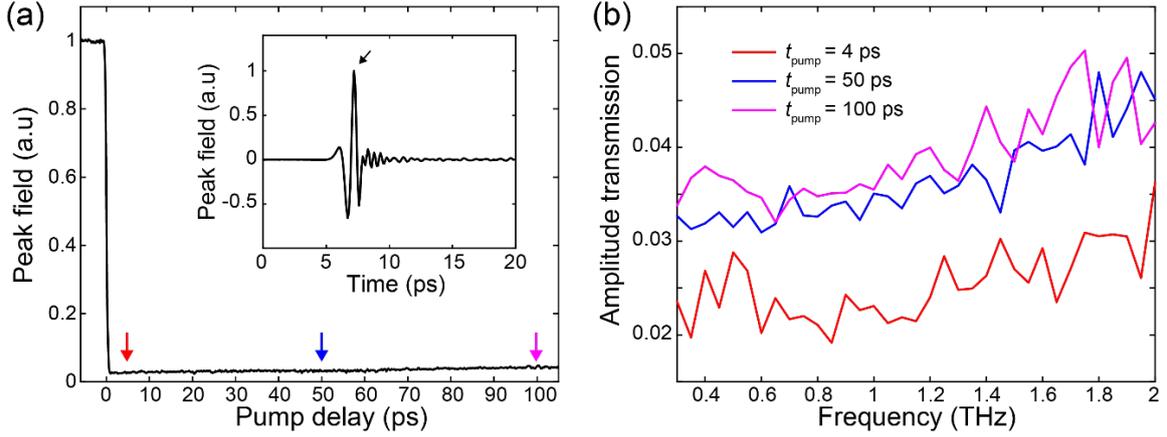

Fig. S1. Optical-pump/THz-probe characterization of the GaAs substrate. (a) The peak field of the transmitted THz pulse through the GaAs substrate is plotted as a function of the time delay between the near-infrared pump and THz probe pulses. The inset shows the representative waveform of a THz probe pulse with its peak field position indicated by an arrow. (b) Amplitude transmission spectra through the temporal mirror measured at $t_{\text{pump}}$ = 4 ps, $t_{\text{pump}}$ = 50 ps, and $t_{\text{pump}}$ = 100 ps.

## 2. Characterization of the static mirror

The spectral characterization of the static mirror (i.e., a thin polyimide film patterned with an array of gold wires) is shown in Fig. S2. The measured transmission through the static mirror, $t(\omega)$, is plotted with crosses in Fig. S2a. From the measured spectra $t(\omega)$, the effective surface conductivity of the static mirror, $\sigma(\omega)$, can be estimated as follows [2]:

$$\sigma(\omega) = \frac{1}{Z_0}\left[\frac{2}{t(\omega)} - 2\right],$$

where $Z_0$ is the impedance of free space. In Fig. S2b, the estimated complex-valued conductivity of the static mirror is plotted as a function of frequency. Additionally, the fitted surface conductivity (red and blue lines) by the Drude model, from which the amplitude transmission $t(\omega)$ can be fitted (a black line in Fig. S2a) are also shown. In the TCMT calculations presented in the main manuscript, the fitted amplitude transmission spectra of the static mirror were used to reproduce the measured resonance-enhanced spectral funneling behavior.

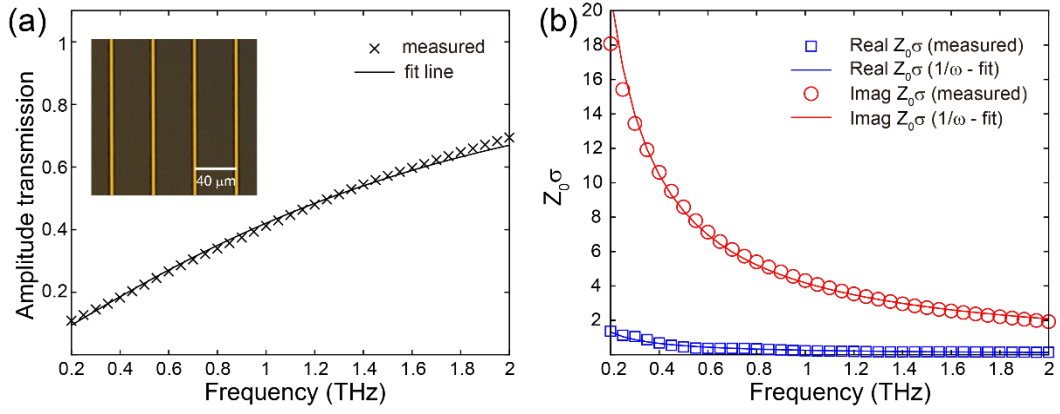

Fig. S2. (a) Amplitude transmission through the static mirror plotted as a function of frequency. The inset shows a microscopic image of the fabricated static mirror. The measured amplitude transmission spectrum is drawn with black crosses, and the Drude model fitting is drawn with a black line. (b) Complex-valued surface conductivity of the static mirror plotted as a function of frequency. Extracted data are drawn with scatters, while the fitted data are drawn with lines.

## 3. Reflection properties of static Fabry–Perot resonators

As the photoconductive layer formed on the surface of the GaAs substrate and the static mirror are much thinner than the wavelength of the input THz pulse, we can apply the following boundary condition to estimate the amplitude reflection spectra from the measured amplitude transmission spectra of the temporal and static mirrors [2]:

$$\widetilde{E}_i(\omega) + \widetilde{E}_r(\omega) - \widetilde{E}_t(\omega) = 0,$$

where $\widetilde{E}_i(\omega)$, $\widetilde{E}_r(\omega)$, and $\widetilde{E}_t(\omega)$ are the spectral amplitudes of the incident, reflected and transmitted THz pulses, respectively. The estimated amplitude reflection spectra, $\widetilde{E}_r(\omega)/\widetilde{E}_i(\omega)$, for the temporal and static mirrors are plotted in Fig. S3.

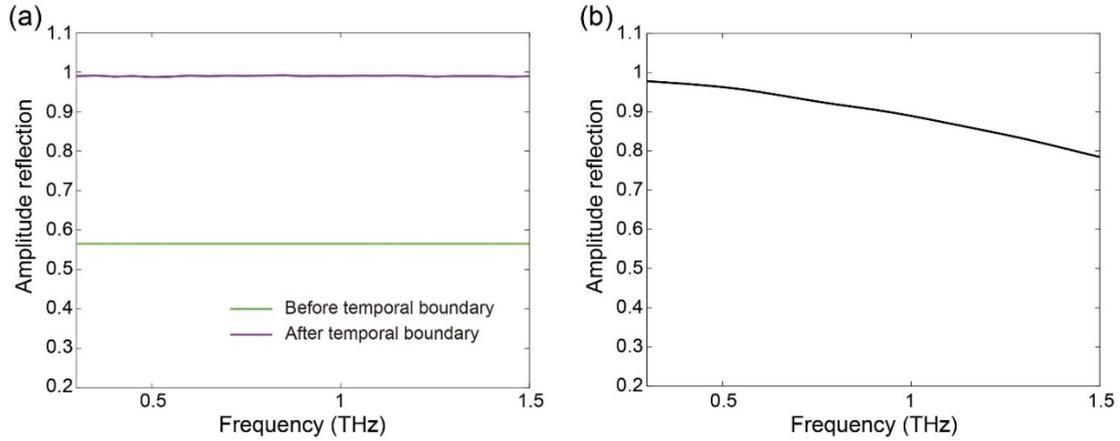

Fig. S3. (a) Estimated amplitude reflection spectra of the temporal mirror. The amplitude reflection spectrum (green line) is obtained by assuming a situation prior to the temporal boundary, while the spectrum (purple line) is obtained by assuming a situation after the temporal boundary. (b) Estimated amplitude reflection spectrum of the static mirror.

With the amplitude reflection spectra of constituting mirrors, those of FP resonators with cavity lengths of 250 μm (Fig. S4a-S4c) and 900 μm (Fig. S4d-S4e) can be estimated by considering multiple interferences [3]. In these calculations, the GaAs substrate was assumed to be semi-infinite (i.e., a half-space filled with GaAs). Here, for simplicity, the incident and reflected THz pulses were assumed to propagate in the GaAs substrate. As seen in Fig. S4a and d, the amplitude reflection spectra of the FP resonators prior to the temporal boundary are characterized by low-Q Fabry–Perot resonances, especially due to the low amplitude reflection of the temporal mirror (see Fig. S3a). On the other hand, after the temporal boundary, the incident THz pulse can barely be coupled to the FP resonator and is almost totally reflected (see Fig. S4a and d). The waveforms of the reflected THz pulses can also be calculated by performing an inverse Fourier transform of the product of the amplitude reflection spectrum and the incident THz pulse spectrum. Fig. S4b and d show that the waveforms reflected from the FP resonators prior to the temporal boundary are characterized by the interference of multiple pulse trains. In contrast, the waveforms reflected from the FP resonators posterior to the temporal boundary are similar to those almost totally reflected as predicted (Fig. S4c and d).

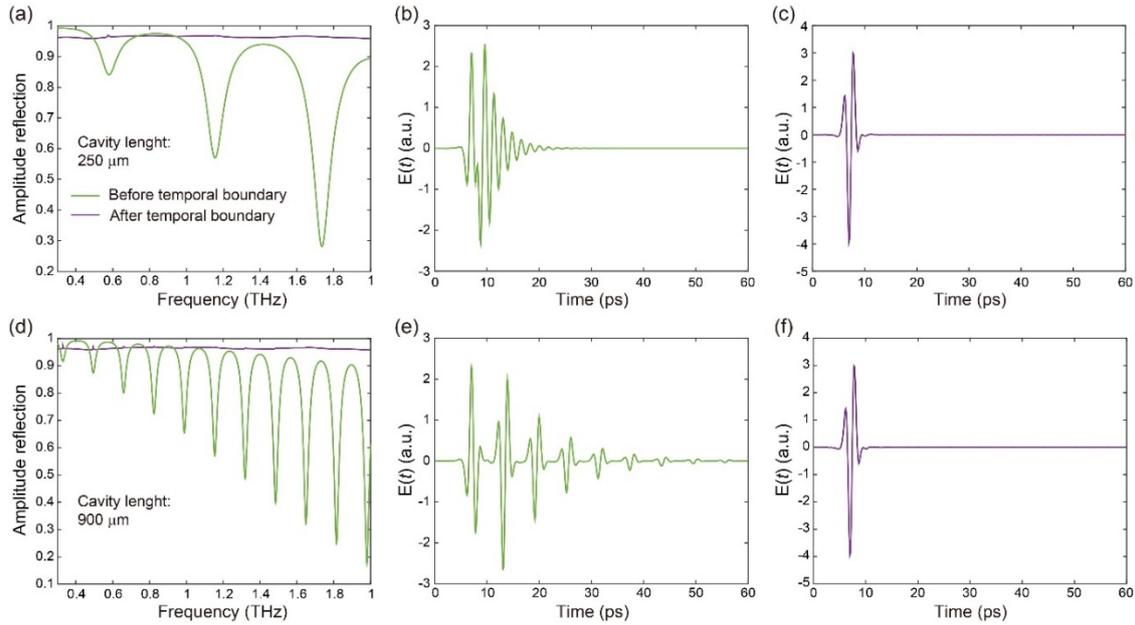

Fig. S4. (a-c) Calculated reflection properties of the Fabry–Perot resonator with a cavity length of 250 μm. (a) Amplitude reflection spectra from the static Fabry–Perot resonators. The green line denotes the amplitude reflection spectrum prior to the temporal boundary, while the purple line denotes the spectrum posterior to the temporal boundary. (b) Calculated waveform reflected from the static Fabry–Perot resonator prior to the temporal boundary. (c) Calculated waveform reflected from the static Fabry–Perot resonator after the temporal boundary. (d-f) Calculated reflection properties of the Fabry–Perot resonator with a cavity length of 900 μm.